\documentclass[apjl]{emulateapj}
\usepackage{epsfig,amsmath,epsfig,color}
\usepackage{natbib}
\newcommand{\fsky}{f_{\rm sky}}
\newcommand{\nhat}{{\bf \hat n}}

\newcommand{\dbestbold}{{\bf d_{best}}}

\newcommand{\dobs}{{\bf d}_{\rm obs}}
\newcommand{\dkin}{{\bf d}_{\rm kin}}
\newcommand{\dlocal}{{\bf d}_{\rm local}}
\newcommand{\Fmarg}{F^{\rm marg}_{(3\times3)}}
\newcommand{\Ngal}{N_{\rm gal}}
\newcommand{\Mpar}{M_{\rm par}}
\newcommand{\lmax}{\ell_{\rm max}}
\newcommand{\zmed}{z_{\rm med}}

\def\kms{\,{\rm km\,s^{-1}}}

\begin{document}

\title{Kinematic dipole detection with galaxy surveys: forecasts and requirements}

\author{Mijin Yoon$^{1}$ and Dragan Huterer$^{1,2,3}$}

\affil{$^{1}$Department of Physics, University of Michigan, 450 Church St, Ann Arbor, MI 48109-1040}
\affil{$^{2}$Max-Planck-Institut for Astrophysics, Karl-Schwarzschild-Str.\ 1, 85741 Garching, Germany}
\affil{$^{3}$Excellence Cluster Universe, Boltzmannstrasse 2, D-85748 Garching,
Germany}

\begin{abstract}
Upcoming or future deep galaxy samples with wide sky coverage can provide
independent measurement of the kinematic dipole -- our motion relative to the
rest frame defined by the large-scale structure. Such a measurement would
present an important test of the standard cosmological model, as the standard
model predicts the galaxy measurement should precisely agree with the existing
precise measurements made using the CMB.  However, the required statistical
precision to measure the kinematic dipole typically makes the measurement
susceptible to bias from the presence of the local-structure-induced dipole
contamination. In order to minimize the latter, a sufficiently deep survey is
required.
We forecast both the statistical error and the systematic bias in the
kinematic dipole measurements. We find that a survey covering $\sim 75\%$ of
the sky in both hemispheres and having $\sim 30$ million galaxies can detect the
kinematic dipole at $5\sigma$, while its median redshift should be at least
$\zmed\sim 0.75$ for
negligible bias from the local structure.

\end{abstract}

\maketitle

\section{Introduction} \label{sec:intro}

Measurements of the motion of our Solar System through the cosmic microwave
background (CMB) rest frame represent one of the early successes of precision
cosmology.  This so-called kinematic dipole corresponds to a velocity of $(369
\pm 0.9)\kms$ in the direction $(l,b) = (263.99^\circ \pm 0.14 ^\circ,
48.26^\circ\pm 0.03^\circ)$ \citep{Hinshaw:2008kr}. The kinematic dipole has
even been detected (though not {\it as} precisely measured) by observing the
relativistic aberration in the CMB anisotropy that it causes, which is
detected via the coupling of high CMB multipoles in Planck
\citep{Aghanim:2013suk}.

Independently, the past few decades have seen significant progress in
measuring the dipole in the distribution of extragalactic sources.  The
contribution of our motion through the large-scale structure (LSS) rest frame
-- the kinematic dipole -- also leads to relativistic aberration, this
time of galaxies or other observed LSS sources. We define the dipole amplitude
via the amount of its ``bunching up'' of galaxies in the direction of the
dipole
\begin{eqnarray}
\label{eq:dN_over_N}
\frac{\delta N({\bf \hat n})}{\bar N}
= A\, \bf \hat d \cdot \bf \hat n + \epsilon(\hat n), 
\end{eqnarray}
where $N$ is the galaxy number in an arbitrary direction ${\bf \hat n}$, ${\bf
  \hat d}$ is the dipole direction, and $\epsilon$ is random noise. The
dipole amplitude $A$ is approximately (but not exactly) equal to our velocity through
the LSS rest frame in units of the speed of light; the precise relation is
given in the following section.

However, the dominant contribution to the LSS dipole is typically not our
motion through the LSS rest frame, but rather the fluctuations in structure
due to the finite depth of the survey.  The dipole component of the latter --
the so-called ``local-structure dipole'' in the nomenclature of
\citet*{Gibelyou2012} -- has amplitude $A\sim 0.1$ for shallow surveys
extending to $z_{\rm max}\sim 0.1$, but is significantly smaller for deeper
surveys. The local-structure dipole is the dominant signal at multipole
$\ell=1$ in all extant LSS surveys. It has been measured and reported either
explicitly
\citep{1998MNRAS.297,blake2002detection,Hirata2009,Gibelyou2012,Fernandez-Cobos:2013fda,Rubart:2013tx,Yoon2014,2014JCAP...10..070A,2015MNRAS.449..670A},
or as part of the angular power spectrum measurements. No LSS survey completed to date
therefore had a chance to separate the small kinematic signal from the larger
local-structure dipole contamination due to insufficient depth and sky
coverage. This will change drastically with the new generation of wide, deep
surveys.

Standard theory based on the adiabatic initial perturbations predicts that the
kinematic dipole measured by the LSS should agree with the one measured by the
CMB.  Detection of an anomalously large (or small) dipole or the disagreement
of its direction from that of the CMB dipole could indicate new physics: for
example, the presence of superhorizon fluctuations in the presence of
isocurvature fluctuations \citep{turner1991tilted,zibin2008gauging,
  itoh2010dipole,erickcek2008superhorizon}.  Clearly, a kinematic dipole
detection and measurement represent an important and fundamental consistency
test of the standard cosmological model.

\section{Methodology} \label{sec:method}

\subsection{Theoretical signal}

The expected LSS kinematic dipole signal amplitude is given by \citep{burles2006detecting,itoh2010dipole} 
\begin{equation}
A = 2 \tilde \beta = 2 [1 + 1.25x(1-p)] \beta 
\label{eq:A_th}
\end{equation}
where $\beta=v/c=0.00123$ (assuming the CMB dipole). The contribution $2\beta$
comes from relativistic aberration, while the correction $[1 + 1.25x(1-p)]$
corresponds to the Doppler effect; here $x$ is the faint-end slope of the
source counts, $x\equiv d\log_{10} [n(m < m_{\rm lim})]/d m_{\rm lim}$, and $p$
is the logarithmic slope of the intrinsic flux density power-law, $S_{\rm
  rest}(\nu) \propto \nu^p$.

Clearly, the parameters $x$ and $p$ depend on the population of sources
selected by the survey, and on any population drifts as a function of
magnitude. We now estimate these parameters -- note also that we only need the
quantity $A$ to set our fiducial model, so very precise values of the
population parameters are not crucial for this paper. \citet{Marchesini_2012}
find that the faint end of the V-band galaxy luminosity function does not vary
much over the redshift range $0.4\leq z\leq 4$ and is equal to, in our
notation, $x=0.11\pm 0.02$. Moreover, for optical sources the flux
  density slope $p$ varies significantly with the age of the source, but in
  the infrared it is more consistent, with measurements indicating $p\sim 0$
  \citep{2004ApJ...602..565W,9780521857932}. Here we adopt $p=0$.  Applying
all these values to Eq.~(\ref{eq:A_th}), we get
\begin{equation}
  A \simeq 0.0028
  \quad \mbox{(expectation from CMB)}.
\label{eq:A_fid_value}
\end{equation}
While the actual value of the kinematic dipole is of course unknown prior to
the measurement, standard cosmology theory predicts it takes this value, plus
or minus O(20\%) changes depending on the source population
selected. We adopt Eq.~(\ref{eq:A_fid_value}) as the fiducial amplitude.

The fiducial direction we adopt is the one of the best-fit CMB dipole, $(l, b)
= (263.99^\circ, 48.26^\circ)$. Note, however, that the results may vary
depending on the relative orientation between the actual dipole direction and
the coverage of the observed sky. Finally, note that bias (of the galaxy
clustering relative to the dark matter field) enters into the contamination of
the kinematic dipole measurements, but not the signal.  The former quantity --
the local-structure dipole -- is linearly proportional to the bias
$b$. Therefore, the bigger the bias, the more contamination the
local-structure dipole provides for measurements of the kinematic effect. In
this work we assume bias of $b=1$.  Note that the kinematic signal itself,
being due to our velocity through the LSS rest frame, is independent of bias.

\subsection{Statistical error}\label{sec:stat_error}

Rewriting Eq.~(\ref{eq:dN_over_N}) somewhat, the modulation in the number of
sources is given at each direction ${\bf\hat n}$ can be written as
\begin{equation}
\label{eq:Ti}
\frac{\delta N(\hat n)}{\bar N} = {\bf x} \cdot {\bf  T(\hat n)}
+ \epsilon(\hat n), 
\end{equation}
where ${\bf x} = (d_x, d_y, d_z, k_1, ..., k_M)$ is the vector of the three
dipole component coefficients in the three spatial coordinates, plus
coefficients corresponding to other multipoles (one for the monopole, five for
that many components of the quadrupole, etc), as well as any desired
systematic templates. The vector ${\bf T(\hat n)} = (n_x, n_y, n_z, t_1(\hat
n), ..., t_M(\hat n))$, contains the three dipole unit vectors (with $n_x^2 +
n_y^2 + n_z^2 =1$), plus $M$ additional spatial patters for all templates
included. Note that the choice of the fiducial values of the non-dipole
template coefficients $k_i$ is arbitrary, since we will fully marginalize over
each of these, effectively allowing $k_i$ to vary from zero to plus
infinity. The optimal estimate of {\bf x} is given by ${\bf \hat x} = F^{-1}
g$ \citep{Hirata2009}, where the components of the vector $g$ are $g_i = \int
T_i(\hat n) \delta N^{\Omega}(\hat n) d^2 \hat n$ and the best-fit dipole
$\dbestbold$ is given by the first three elements of ${\bf x}$. Here the
Fisher matrix $F$ is given by
\begin{equation}
F_{ij} =\bar N^{\Omega} \int T_i(\hat n) T_j(\hat n) d^2 \hat n,
\end{equation}
where $N^{\Omega} \equiv dN/d\Omega$ is the number of galaxies per steradian
and $\Omega$ is a solid angle. Note that the Fisher information is
proportional to the number of sources, and unrelated to the depth of the
survey. It is therefore the number of sources, together with the sky cut (not
just the fraction of the sky observed $\fsky$ but also the shape of the
observed region relative to the multipoles that need to be extracted) that
fully determines the statistical error in the various templates including the dipole.


The Fisher matrix contains information about how well the three Cartesian dipole
components, as well as the multipole moments of all other components, can be
measured in a given survey. Our parameter space has a total of
$\Mpar=(\lmax+1)^2$ parameters, where $\lmax$ is the maximum multipole
included to generate the templates (see below for more on the choice of
$\lmax$). With this Fisher matrix in hand, we then marginalize over the
$M\equiv \Mpar-3$ non-dipole parameters, using standard Fisher techniques, to
get the $3\times 3$ Fisher matrix describing the final inverse covariance
matrix for the dipole components.  Finally, we perform a basis change,
converting from Cartesian coordinates $\{d_x, d_y, d_z\}$ to spherical
coordinates $\{A, \theta, \phi\}$ (where $A$ is the amplitude of dipole), by
using a Jacobian transformation to obtain the desired $3\times 3$
Fisher matrix in the latter space, $\Fmarg$. The forecasted error on $A$ is
then given in terms of this matrix as
\begin{eqnarray}
\sigma(A)= \sqrt{[(\Fmarg)^{-1}]_{AA}}.
\end{eqnarray}

In a realistic survey with partial-sky coverage, the presence of other
multipoles (monopole, quadrupole, etc) will be degenerate with the dipole,
degrading the accuracy in determining the latter.  We have extensively tested
for this degradation, in particular with respect to how many multipoles need
to be kept -- that is, what value of $\lmax$ (and therefore $M$) to adopt. We
explicitly found that the prior information on the ``nuisance'' $C_\ell$,
corresponding to how well they can be (and are being) independently measured,
is of key value: once the prior information on the $C_\ell$ -- corresponding
to cosmic variance plus measurement error -- is added, very high multipoles
are not degenerate with the dipole. Our tests show that keeping all multipoles
out to $\lmax=10$ is sufficient for the dipole error to fully converge. We
also experimented with adding additional individual templates $t_i(\nhat)$
corresponding to actual sky systematics and with modified coverage
(corresponding to e.g.\ dust mask around the Galactic plane), but found that
these lead to negligible changes in the results. Moreover, we envisage a
situation where the maps have already been largely cleaned of stars by the
judicious choice of color cuts prior to the dipole search analysis. For these two reasons, we choose not to include any
additional systematic templates in the analysis.

\subsection{Systematic bias}

\begin{figure}[t]
\includegraphics[width=0.48\textwidth]{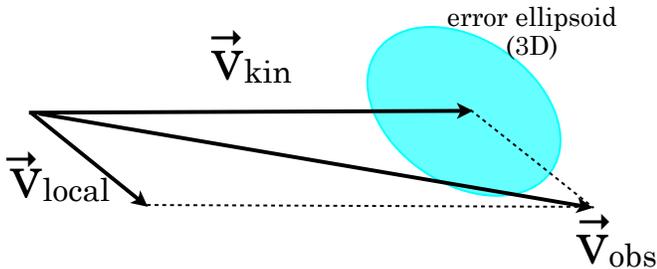}
\caption{ Sketch of the problem at hand: we would like to measure the
  kinematic dipole $\vec{v}_{\rm kin}$, whose error (represented by a cyan
  ellipse) can be calculated given the number of extragalactic objects and
  the sky coverage. The LSS local dipole, $\vec{v}_{\rm local}$,
  provides a bias in this measurement. For a survey deep enough (and depending
  somewhat on the direction of its $\vec{v}_{\rm local}$), bias in the measurement of
  $\vec{v}_{\rm kin}$ will be smaller than the statistical error.  }
\label{fig:sketch}
\end{figure}

The local-structure dipole $\dlocal$ will also provide a contribution to the
kinematic signal $\dkin$ that we seek to measure. The observed dipole in any
survey will be the sum of the two contributions:
\begin{equation}
\dobs = \dkin +\dlocal.
\end{equation} 
Without any loss of generality, we consider the kinematic dipole as the
fiducial signal in the map, whose errors are therefore given by the Fisher
matrix worked out above. We consider the local-structure dipole to represent
the contaminant whose magnitude, ideally, should be such that the resulting
observed dipole $\dobs$ is still within the error ellipsoid around the
kinematic dipole direction and amplitude. This is illustrated in Fig. 1.

We now quantify the systematic bias, due to the local structure, relative to
statistical error in the measurements of the kinematic dipole.  First note
that we are in possession of the (statistical) inverse covariance matrix for
measurements of the kinematic dipole, $\Fmarg$, which is already fully
marginalized over other templates. The quantity
\begin{eqnarray}
  \nonumber
\Delta \chi^2(\dlocal) &=&
(\dobs -\dkin) ^T \Fmarg (\dobs - \dkin) \\[0.2cm]
&=&  \dlocal^T\Fmarg \dlocal
\label{eq:chisq}
\end{eqnarray}
then represents ``(bias/error)$^2$'' in the kinematic dipole measurement due
to the presence of the local-structure contamination. This chi squared depends
quadratically on the expected local-structure dipole, and is therefore
expected to sharply drop with deeper surveys which have a lower $|\dlocal|$,
as we find in the next section. With three parameters, requiring 68\%
confidence level departure implies $\Delta \chi^2=3.5$. We therefore require
that, for a given survey, the local-structure dipole magnitude and direction
are such that the value in Eq.~(\ref{eq:chisq}) is smaller than this value.

\section{Results}

\begin{figure*}[t]
\includegraphics[width=0.5\textwidth]{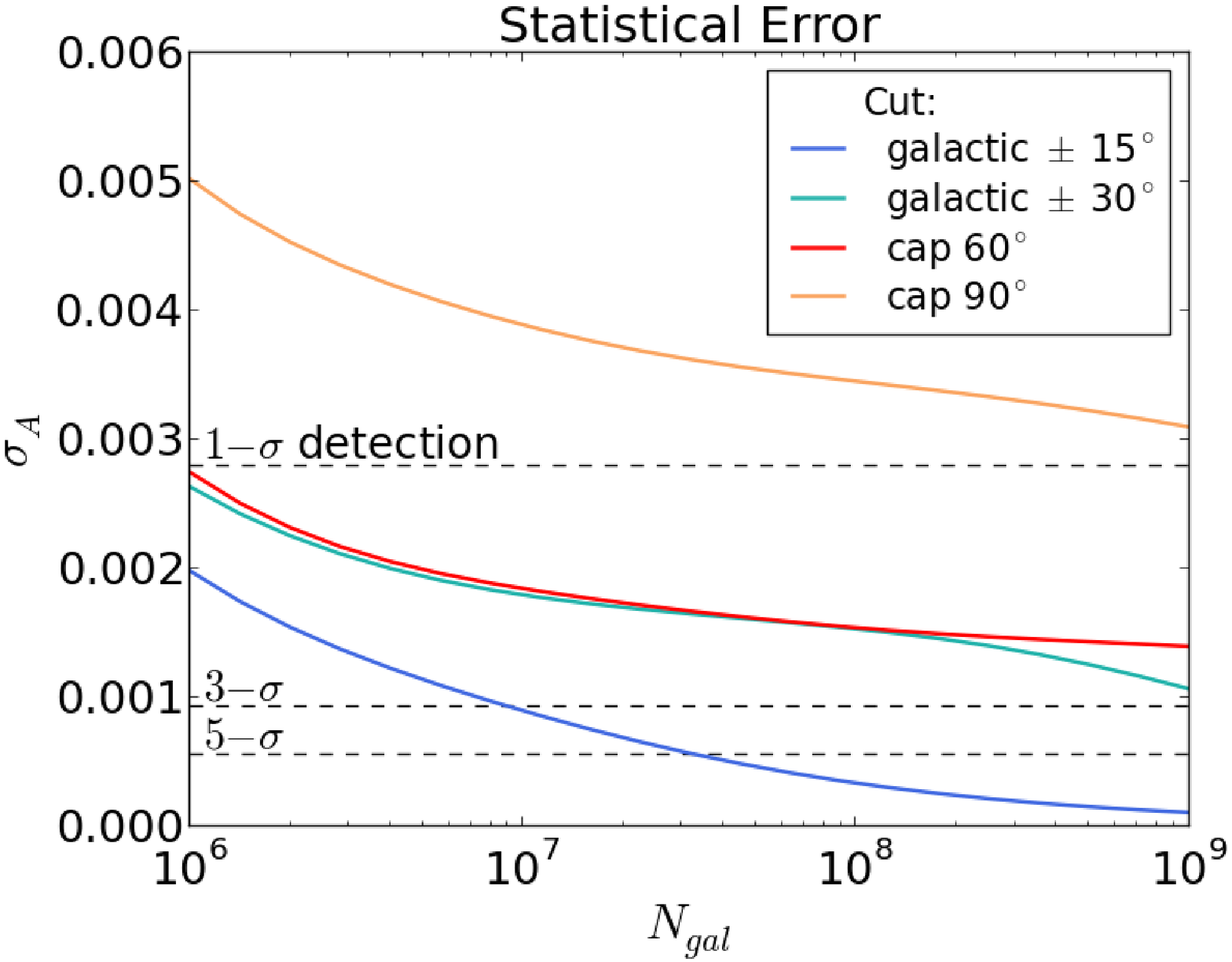}
\includegraphics[width=0.5\textwidth]{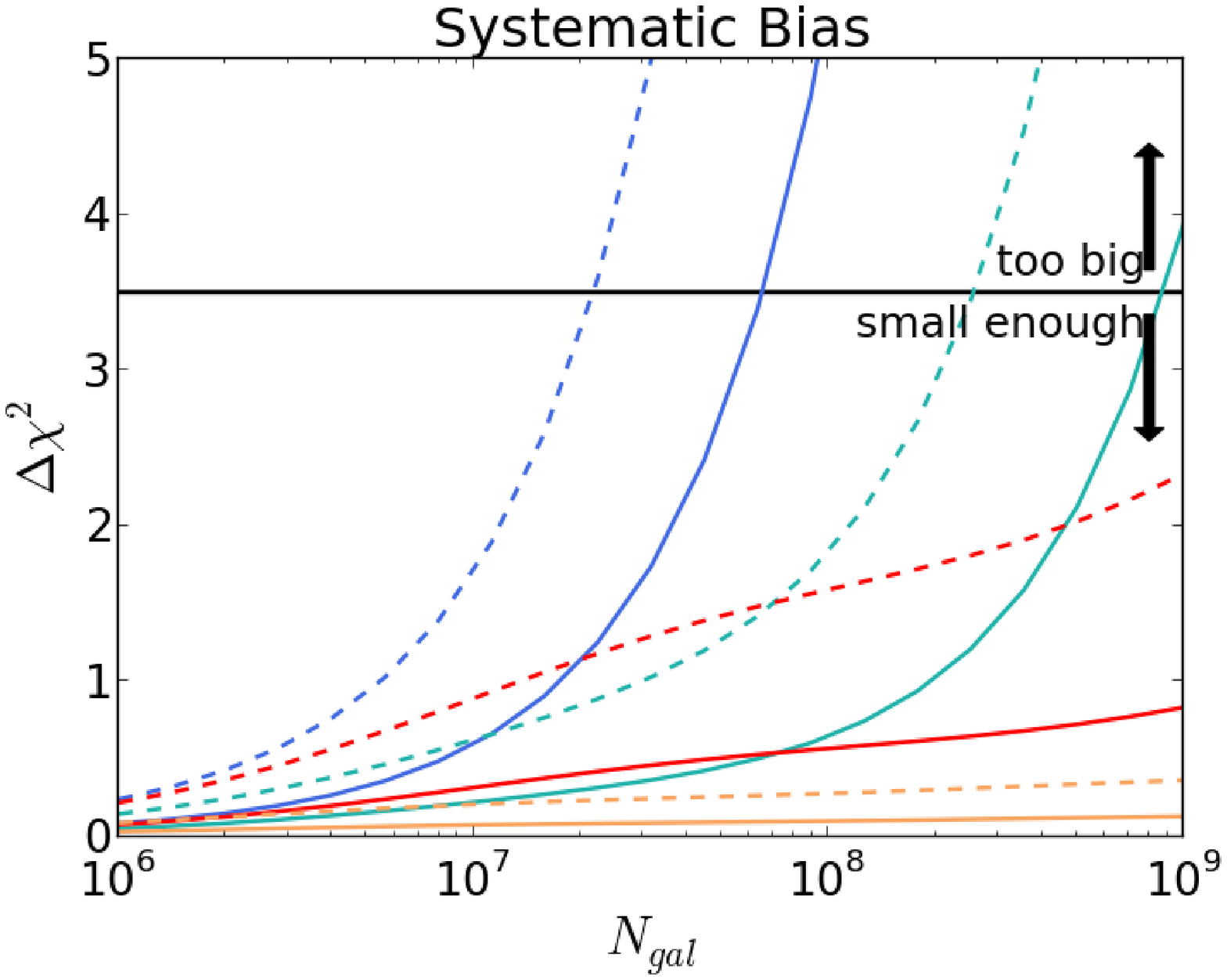}
\caption{ Left panel: Statistical error in the dipole amplitude, marginalized
  over direction and other multipoles that are coupled to the dipole, as a
  function of the number of galaxies in a survey. The top horizontal dashed
  line shows the amplitude expected based on the CMB dipole measurements
  ($A=0.0028$), and is the fiducial value in this work. The two dashed
  horizontal lines show the $3\sigma$ and $5\sigma$ detection of dipole with
  the fiducial amplitude. Right panel: $\Delta \chi^2$, defined in
  Eq.~(\ref{eq:chisq}), corresponding to the bias from the local-structure
  dipole, as a function of the number of objects $\Ngal$. For a fixed
  amplitude of $\dlocal$ the $\Delta\chi^2$ still depends on the direction of
  this vector; here we show the value averaged over all directions of
  $\dlocal$. Solid lines show cases when the median galaxy redshift is $\zmed
  = 1.0$, while dashed lines are for $\zmed=0.75$. The legend colors are the
  same in the left panel.  }
\label{fig:amp_stat_error}
\end{figure*}

For a fixed sky cut and number of sources in the survey, we first calculate
the error on the amplitude of the dipole $\sigma(A)$. In the left panel of
Fig.~\ref{fig:amp_stat_error} we show errors as a function of the number of
galaxies in a survey. As previously noted, this statistical error does not
depend on the depth of the survey, but does depend on both $\fsky$ and the
shape of the sky coverage. Here we show results for an isolatitude cut around
the equator of $\pm 15\deg$, and $\pm 30\deg$ and isolatitude cap-shaped cuts
of $90\deg$ (i.e.\ half the sky removed) and $60\deg$ (i.e.\ leaving out a
circular region around a pole). Note that the Galactic $\pm 15\deg$ cut and
the cap cut of $60\deg$ both have $\fsky=0.75$, while the Galactic $\pm
30\deg$ and the cap $90\deg$ cuts both have $\fsky=0.5$.  The results will
also depend on the fiducial amplitude of the dipole, and here and throughout
we assume the CMB-predicted value of $A=0.0028$. Even for a fixed $\fsky$ of
the survey, the cut geometry clearly matters, and the Galactic-cut cases have
a smaller error in the dipole amplitude due to symmetrical covering of the two
hemispheres. For Galactic $\pm 15\deg$ case, 3-$\sigma$ and 5-$\sigma$
detections are easily achievable, requiring only $\Ngal=9\times10^6$ and
$3\times10^7$ objects, respectively.  Note that if the actual
  direction of the LSS kinematic dipole deviates from the assumed dipole
  direction (the CMB direction), the result changes. For a 5-sigma detection
  and the same  $\pm 15\deg$  isolatitude cut, a dipole pointing along toward a Galactic
  pole, which is the best-case scenario, only requires 8 million objects; if
  instead the dipole points toward the Galactic plane, then
  70 million objects are required. For the
Galactic $\pm 30\deg$ cut, the 3-$\sigma$ detection is more challenging since
it requires having over $\Ngal=10^9$ sources. The cap $60\deg$ cut mostly
follows the trend of the Galactic $\pm 30$ case. Lastly, the $90\deg$ cap cut
cannot detect the signal even at the 1-sigma level and with $\Ngal=10^9$. We
conclude that dual-hemisphere sky coverage is crucial in the ability of the
survey -- or a combined collection of surveys -- to detect the kinematic
dipole.

The right panel of Fig.~\ref{fig:amp_stat_error} shows the systematic bias in
the dipole measurement due to the presence of the local-structure
contamination, showing the quantity defined in Eq.~(\ref{eq:chisq}). Because
$\dlocal$ has an a-priori unknown direction and its amplitude changes
according to the depth of the survey, the systematic error is a function of
direction of $\dlocal$ and the depth of the survey. Therefore, we choose to plot 
 $\Delta \chi^2$ averaged over all directions of $\dlocal$.

To calculate the amplitude of $\dlocal$, we model the radial distribution of
objects as $n(z) =z^2/ (2z_0^{3}) \exp(-z/z_0)$ \citep{Huterer:2001yu}, where
the parameter $z_0$ is related to the median redshift as $z_0 =\zmed/2.674$. A
deeper survey (larger $z_0$) has a smaller local-structure dipole. Note that
one could additionally cut out low-redshift objects in order to further reduce
the contamination from the nearby structures, as well as the star-galaxy
confusion; we have not done that here.


The dashed lines in the right panel of Fig.~\ref{fig:amp_stat_error} represent
the cases when $\zmed = 0.75$ and the solid ones are when $\zmed = 1.0$. 
Since $\Delta \chi^2$ is inversely proportional to the statistical error
squared, the best cases in the left panel of Fig.~\ref{fig:amp_stat_error}
have a larger bias in the right panel.  In particular, the more galaxies
the survey has, the more it is susceptible to systematic bias (for a fixed
depth and thus $|\dlocal|$). For example, a survey with $\pm 15\deg$ Galactic
cut with 30 million sources can detect the kinematic dipole at
5-$\sigma$, but needs to have a median redshift of at least $\zmed=0.75$ in
order for this not to be excessively biased due to local structures.

On the whole, Fig.~\ref{fig:amp_stat_error} indicates that the convincing
detection of the kinematic dipole expected given the CMB measurements is
entirely within reach of future surveys, as long as those surveys have good
coverage over both hemispheres and, given the source density, are deep enough
not to be biased by the local-structure dipole. All requirements can be
straightforwardly satisfied by surveys like some combination of LSST
\citep{Ivezic:2008fe}, Euclid \citep{Laureijs:2011gra} and DESI
\citep{Levi:2013gra} and, especially, by deep, all-sky surveys with good
redshift information such as SPHEREX \citep{Dore:2014cca}.

Finally, we have also calculated the statistical error in the direction of the
kinematic dipole, based on the fiducial amplitude we had adopted. The
direction's error is generally rather large, e.g.\ an area of about $\simeq
10\deg$ in radius for $\Ngal=10^8$ and the Galactic $\pm 15\deg$
cut. Nevertheless, a combination of the kinematic dipole's amplitude {\it
  and} direction that roughly match the CMB dipole would present a convincing
confirmation of the standard assumption. One could further carry out detailed
forecasts of what various findings could rule out the null hypothesis; we
leave that for future work.

\section{Conclusions}

We have studied the prospects for measuring the kinematic dipole -- our motion
through the LSS rest frame -- as revealed by the relativistic aberration of tracers of the large-scale structure. The standard theory
predicts that the kinematic dipole should agree with the CMB dipole, but this
expectation could be violated due to a number of reasons. Therefore, verifying
the standard expectation is an important null test in cosmology. The 
challenge comes from the fact that the dipole amplitude is small ($A\sim
0.003$), and easily contaminated by the intrinsic clustering of galaxies (the
``local-structure dipole'').

A successful measurement of the kinematic dipole therefore has two
qualitatively different requirements: the survey should cover most of the sky
and have enough objects to have sufficient signal-to-noise to detect the
aberration signature of the dipole, but it should also be deep enough, so that
the local-structure dipole contamination is sufficiently small. The two
requirements are displayed in the two panels of Fig.~\ref{fig:amp_stat_error}
respectively. For a 5-$\sigma$ detection, a survey covering $\gtrsim 75\%$ of
the sky in both hemispheres (our ``Galactic $\pm 15\deg$ cut'' case), with
$\Ngal\gtrsim 30$ million galaxies, is required. For a negligible bias, this
same survey should have median redshift greater than about $0.75$ or
higher, with increasing depth requirements as $\Ngal$ increases.

Fortunately these requirements can be satisfied by upcoming
surveys, including DESI, Euclid, and LSST if they are combined, and
potentially with SPHEREX alone. Even current all-sky surveys such as WISE
(Wide-field Infrared Survey Explorer, \cite{2010AJ}) are not out of the
question, provided a sufficiently deep sample can be selected photometrically;
current WISE samples have typical galaxy redshifts $\zmed\simeq 0.2$
\citep{2014ApJS..210....9B} and are not yet deep enough to measure the
kinematic dipole.

\section*{Acknowledgments}
Our work has been supported by NSF under contract AST-0807564 and DOE under
contract DE-FG02-95ER40899, and also by the DFG cluster of excellence ``Origin
and Structure of the Universe'' (\url{www.universe-cluster.de}). We thank
Cameron Gibelyou for comments on the manuscript.

\bibliographystyle{apj}
\bibliography{kinem_dipole_forecast}

\begin{thebibliography}{}
\expandafter\ifx\csname natexlab\endcsname\relax\def\natexlab#1{#1}\fi

\bibitem[{Aghanim {et~al.}(2014)}]{Aghanim:2013suk}
Aghanim, N., {et~al.} 2014, \aap, 571, A27

\bibitem[{{Alonso} {et~al.}(2015){Alonso}, {Salvador}, {S{\'a}nchez}, \&
  et~al.}]{2015MNRAS.449..670A}
{Alonso}, D., {Salvador}, A.~I., {S{\'a}nchez}, F.~J., \& et~al. 2015, \mnras,
  449, 670

\bibitem[{{Appleby} \& {Shafieloo}(2014)}]{2014JCAP...10..070A}
{Appleby}, S., \& {Shafieloo}, A. 2014, JCAP, 10, 70

\bibitem[{{Baleisis} {et~al.}(1998){Baleisis}, {Lahav}, {Loan}, \&
  {Wall}}]{1998MNRAS.297}
{Baleisis}, A., {Lahav}, O., {Loan}, A.~J., \& {Wall}, J.~V. 1998, \mnras, 297,
  545

\bibitem[{{Bilicki} {et~al.}(2014){Bilicki}, {Jarrett}, {Peacock}, {Cluver}, \&
  {Steward}}]{2014ApJS..210....9B}
{Bilicki}, M., {Jarrett}, T.~H., {Peacock}, J.~A., {Cluver}, M.~E., \&
  {Steward}, L. 2014, \apjs, 210, 9

\bibitem[{Blake \& Wall(2002)}]{blake2002detection}
Blake, C., \& Wall, J. 2002, Natur, 416, 150

\bibitem[{Burles \& Rappaport(2006)}]{burles2006detecting}
Burles, S., \& Rappaport, S. 2006, \apjl, 641, L1

\bibitem[{Dor\'{e} {et~al.}(2014)}]{Dore:2014cca}
Dor\'{e}, O., {et~al.} 2014, arXiv:1412.4872

\bibitem[{Erickcek {et~al.}(2008)Erickcek, Carroll, \&
  Kamionkowski}]{erickcek2008superhorizon}
Erickcek, A., Carroll, S., \& Kamionkowski, M. 2008, \prd, 78, 083012

\bibitem[{Fern\'{a}ndez-Cobos {et~al.}(2014)Fern\'{a}ndez-Cobos, Vielva,
  Pietrobon, \& et~al.}]{Fernandez-Cobos:2013fda}
Fern\'{a}ndez-Cobos, R., Vielva, P., Pietrobon, D., \& et~al. 2014, \mnras,
  441, 2392

\bibitem[{Gibelyou \& Huterer(2012)}]{Gibelyou2012}
Gibelyou, C., \& Huterer, D. 2012, \mnras, 427, 1994

\bibitem[{Hinshaw {et~al.}(2009)}]{Hinshaw:2008kr}
Hinshaw, G., {et~al.} 2009, \apjs, 180, 225

\bibitem[{Hirata(2009)}]{Hirata2009}
Hirata, C.~M. 2009, JCAP, 0909, 011

\bibitem[{Huterer(2002)}]{Huterer:2001yu}
Huterer, D. 2002, \prd, 65, 063001

\bibitem[{Itoh {et~al.}(2010)Itoh, Yahata, \& Takada}]{itoh2010dipole}
Itoh, Y., Yahata, K., \& Takada, M. 2010, \prd, 82, 043530

\bibitem[{Ivezic {et~al.}(2008)Ivezic, Tyson, Allsman, Andrew, \&
  Angel}]{Ivezic:2008fe}
Ivezic, Z., Tyson, J.~A., Allsman, R., Andrew, J., \& Angel, R. 2008,
  arXiv:0805.2366

\bibitem[{Laureijs {et~al.}(2011)}]{Laureijs:2011gra}
Laureijs, R., {et~al.} 2011, arXiv:1110.3193

\bibitem[{Levi {et~al.}(2013)}]{Levi:2013gra}
Levi, M., {et~al.} 2013, arXiv:1308.0847

\bibitem[{{Marchesini} {et~al.}(2012){Marchesini}, {Stefanon}, {Brammer}, \&
  {Whitaker}}]{Marchesini_2012}
{Marchesini}, D., {Stefanon}, M., {Brammer}, G.~B., \& {Whitaker}, K.~E. 2012,
  \apj, 748, 126

\bibitem[{Mo {et~al.}(2010)Mo, van~den Bosch, \& White}]{9780521857932}
Mo, H., van~den Bosch, F., \& White, S. 2010, Galaxy Formation and Evolution
  (Cambridge University Press)

\bibitem[{Rubart \& Schwarz(2013)}]{Rubart:2013tx}
Rubart, M., \& Schwarz, D.~J. 2013, \aap, 555, A117

\bibitem[{Turner(1991)}]{turner1991tilted}
Turner, M. 1991, \prd, 44, 3737

\bibitem[{{White} \& {Majumdar}(2004)}]{2004ApJ...602..565W}
{White}, M., \& {Majumdar}, S. 2004, \apj, 602, 565

\bibitem[{{Wright} {et~al.}(2010)}]{2010AJ}
{Wright}, E.~L., {et~al.} 2010, \aj, 140, 1868

\bibitem[{{Yoon} {et~al.}(2014){Yoon}, {Huterer}, {Gibelyou}, {Kov{\'a}cs}, \&
  {Szapudi}}]{Yoon2014}
{Yoon}, M., {Huterer}, D., {Gibelyou}, C., {Kov{\'a}cs}, A., \& {Szapudi}, I.
  2014, \mnras, 445, L60

\bibitem[{Zibin \& Scott(2008)}]{zibin2008gauging}
Zibin, J., \& Scott, D. 2008, \prd, 78, 123529

\end{thebibliography}

\end{document}